# Thermodynamically stable equal-module exchange magnetic classes.

Khisa Sh. Borlakov* and Albert Kh. Borlakov
*North Caucasian State Humanitarian and Technological Academy,
36 Stavropolskaya str., Cherkessk, Russia, 369001*

It is shown that within the framework of application of the general scheme of the Landau theory of phase transitions to the magnetic crystals the equilibrium and stable equal-module exchange structures automatically appears in addition to the other magnetic states. No additional conditions such as the existence of the Andreev-Marchenko spin scalar are needed. Furthermore, they are not constrained by the dimensionality of irreducible representation as it typically takes place for the Andreev-Marchenko-type structures.

## PACS number(s): 61.50.Ah, 64.70.Kb, 71.70.Ej, 75.30.Gw

### 1. Introduction

The magnetic symmetry of magnetically ordered crystals in the exchange approximation is higher than in the case when relativistic interactions are taken into consideration. A group theory method of determining the symmetry of exchange structures was proposed by Andreev and Marchenko [1] and by Bar'yakhtar and Yablonskii [2] on the basis of the condition for the existence of a certain invariant function formed by spin components (spin scalar). The exchange magnetic structures obtained by using this approach are equal-module structures, i.e., the magnitude of spin moment is the same in all magnetic positions of the crystal. Gufan *et al* [3] used a model example to prove the existence of exchange structures with different modules. We propose a method for obtaining exchange magnetic structures which is more general than magnetic classes obtained from the condition for the existence of a spin scalar.

For this purpose, a general approach is required, which would not require any information other than that provided by the magnetic symmetry group of the paramagnetic phase and the population density of magnetic atoms in positions in the space group of the crystal. If such information is available, all possible types of exchange structures for the given crystal can be determined on the basis of the group theory method of determining low-symmetry phases proposed in Refs. [4-6]. We consider the problem of finding the equilibrium exchange classes, according to [7].

### 2. General method of determining exchange magnetic structures

The magnetic symmetry of the paramagnetic phase of a crystal is characterized by the exchange paramagnetic group *M* that can be reduced to a direct product of the space group *G* and the three-dimensional group of spin rotations O(3), i.e., $M = G \times O(3)$ [8]. Each space group is characterized by a set of crystallographic orbits or a regular system of points (RSP) [9]. Any point belonging to a RSP is transformed into a point from the same RSP under the action of symmetry elements of the space group *G*, i.e., RSP is transformed into itself under the action of elements from group *G*. Accordingly, any system of functions denned on RSP is transformed into itself under the action of symmetry elements from *G*, i.e., forms the basis of the irreducible representation (IR) of this group. In addition to the symmetry group, a given crystal is characterized by the diagram of population of various RSP by atoms. Some crystallographic orbits in a magnetic crystal are populated completely or partially by magnetic atoms. Each such orbit can be presented by the spin density function S(r). We expand this function in the basis functions from the IR of the group *M*:

$$\vec{S}(\vec{r}) = \sum_{a=1}^{3}\sum_{i=1}^{n} S_a^i \varphi_i(\vec{r}) \cdot \vec{e}^a = \sum_i \vec{S}_i \varphi_i(\vec{r}), \qquad (1)$$

where $\varphi_i(\vec{r})$ are the basis functions of the IR of the group G, $\vec{e}^a$ is the orthonormal basis in the spin space (the basis of the vectorial IR of the O(3) group), *a* the index labelling unit vectors in the spin space, and *i* the index labelling basis functions of the IR. Thus, the type of magnetic ordering is determined by the set of the mixing coefficients $S_a^i$. The last two factors on the right-hand side of formula (1) form the IR $D = \Gamma \times V$ of the $G \times O(3)$ group, where $\Gamma$ is the IR of the *G* group, whose basis is just formed by functions appearing in (1), and *V* is the vectorial IR of the group of three-dimensional spin rotations. Spin group transformations correspond to transpositions of atoms within a crystallographic orbit. In addition to lattice site coordinates, magnetic atoms are also characterized by spin. However, the direction and magnitude of spin do not change upon a transposition of an atom, i.e., spin rotations and space transformations are carried out independently. This is a formally logical realization of the condition corresponding to the absence of a coupling between the spin subsystem and the lattice, which is executed by relativistic interactions. Thus, the transformation properties of each component of the spin density function are similar to transformation properties of variation of electric charge density. Let us clarify this statement. The variation $\Delta\rho(\vec{r})$ of electric charge density has two components [10]: the variation of the form of the function $\delta\rho(\vec{r})$ and the variation of the function due to a change in the argument:

$$\Delta\rho(\vec{r}) = \delta\rho(\vec{r}) + \frac{\partial\rho}{\partial\vec{r}}\vec{u}(\vec{r}) \; , \tag{2}$$

where $\vec{u}(\vec{r})$ is the vector function of atomic displacements.

The variation of the form of the function describes purely phase transitions associated with atomic ordering. The second term on the right-hand side of (2) characterizes purely displacement-type transitions. The transformation properties of the spin density function components are identical just with the transformation properties of the function $\delta\rho$ describing atomic ordering. The (reducible) representation according to which the function $\delta\rho$ is transformed is called the transposition representation [9]. Thus, the spin density function can be expanded in the basis functions of the IR appearing in the transposition representation [9].

Atomic ordering can be described by a set of scalar basis functions and a set of mixing coefficients $\vec{c} = (c_1, c_2, ..., c_n)$ forming the so-called stationary vector (statvector) [3,6]. In order to describe the exchange magnetic ordering, we need a set of scalar basis functions and three statvectors

$$(S_1^1, S_2^1, ..., S_n^1), (S_1^2, S_2^2, ..., S_n^2), (S_1^3, S_2^3, ..., S_n^3) \; . \tag{3}$$

The statvector c can be regarded a vector of a certain vector space (e-space in the terminology of Gufan [4]) Different symmetry positions of the statvector in this space correspond to different subgroups of the group $G_D \subset G$ of the high-symmetry phase. Each of these subgroups describes the crystal symmetry corresponding to atomic ordering with certain mixing coefficients.

The three statvectors (3) can be conveniently combined into a stationary matrix (statmatrix) of dimensionality $n \times 3$:

$$\hat{S} = \begin{pmatrix} S_1^1, S_2^1, ..., S_n^1 \\ S_1^2, S_2^2, ..., S_n^2 \\ S_1^3, S_2^3, ..., S_n^3 \end{pmatrix} \tag{4}$$

The exchange magnetic ordering symmetry is determined by the least symmetric of the three statvectors appearing in statmatrix (4). If we go over to the spherical coordinates in the spin space, the statmatrix assumes the form

$$\hat{S} = \begin{pmatrix} S_1 \sin\theta_1 \cos\varphi_1; & S_2 \sin\theta_2 \cos\varphi_2; & ...; & S_n \sin\theta_n \cos\varphi_n \\ S_1 \sin\theta_1 \sin\varphi_1; & S_2 \sin\theta_2 \sin\varphi_2; & ...; & S_n \sin\theta_n \sin\varphi_n \\ S_1 \cos\theta_1; & S_2 \cos\theta_{;2} & ...; & S_n \cos\varphi_n \end{pmatrix} \tag{5}$$

In the general case, statmatrix (5) defines an essentially three-dimensional magnetic structure. If, however, one of the spherical angles $\theta_i$ or $\varphi_i$, does not depend on the number *i* of the column, the statmatrix can be reduced to a form with a single zero line by rotation of the spin axes. Such a statmatrix corresponds to two-dimensional exchange structures. If the other spherical angle is also the same for all the column, the rotation of spin axes can nullify two rows of the matrix *S*, and such a matrix corresponds to a collinear antiferromag-netic exchange structure.

A unit IR of the space group G corresponds to the ferromagnetic spin ordering within each crystallographic orbit. Other one-dimensional IR appearing in the transposition representation at the crystal lattice sites occupied by magnetic atoms correspond to collinear antiferro-magnetic exchange structures. Two-dimensional IR appearing in the transposition representation correspond two-dimensional antiferromagnetic structures. Irreducible representations with a dimensionality higher than three and appearing in the transposition representation correspond to essentially three-dimensional antiferromagnetic structures. Obviously, a change in the number of independent parameters in the statmatrix corresponds to a transition from one exchange phase to another as in the case of an atomic ordering.

It should be noted that by the dimensionality of an IR we meant the dimensionality of real-valued IR or the dimensionality of physically-irreducible representations of complex IR.

### 3. Equal-module exchange structures

Let us consider the exchange structures obtained from the condition for the existence of the Andreev-Marchenko spin scalar. Raising both sides of equality (1) to the second power, we obtain

$$S^2(\vec{r}) = \sum_{a,i}\sum_{b,k} S_a^i S_b^k \varphi_i(\vec{r})\varphi_k(\vec{r}) \cdot \vec{e}^a \cdot \vec{e}^b \; . \tag{6}$$

The left-hand side of this equation is a scalar relative to transformations from space group G. Consequently, the right-hand side of this relation must also assume a form invariant to spatial transformations:

$$S^2(\vec{r}) = \sum_a S_a^2 \sum_i |\varphi_i(\vec{r})|^2 \tag{7}$$

A comparison of the right-hand sides of (6) and (7) shows that the mixing coefficients must satisfy the following orthogonality condition:

$$S_a^i S_b^k = S_a^2 \delta_{ik} \delta_{ab} \tag{8}$$

where $\delta_{ik}$ and $\delta_{ab}$ are Kronecker deltas. For the sake of visualization, it is convenient to write relations (1), (6)-(8) in matrix form. We introduce the column vector

$$\vec{\varphi}(\vec{r}) = \begin{pmatrix} \varphi_1(\vec{r}) \\ \varphi_2(\vec{r}) \\ \dots \\ \varphi_n(\vec{r}) \end{pmatrix}$$

and denote the line vector by $\vec{\varphi}^*(\vec{r})$. Then expansion (1) can be written in the matrix form

$$\vec{S}(\vec{r}) = \hat{S} \cdot \vec{\varphi}(\vec{r}),$$

and expression (7) can be written as

$$S^2(\vec{r}) = \vec{\varphi}^*(\vec{r}) \cdot \hat{S}^T \cdot \hat{S} \cdot \vec{\varphi}(\vec{r}),$$

where $S^T$ is the transposed statmatrix. We present the stat-matrix $S$ in the form of the row

$$\hat{S} = (\vec{S}_1, \vec{S}_2, \dots, \vec{S}_n),$$

where the vectors $\vec{S}_i$,- are defined in the three-dimensional spin space. In such a case, the condition (8) of invariance of the square of spin density module assumes the form

$$\hat{S}^T \hat{S} = \begin{pmatrix} \vec{S}_1 \\ \vec{S}_2 \\ \dots \\ \vec{S}_n \end{pmatrix} (\vec{S}_1, \vec{S}_2, \dots, \vec{S}_n) = \vec{S}^2 \cdot \hat{E} \tag{9}$$

This relation (9) indicates that the $n$ vectors $S_i$ forming the statmatrix are mutually orthogonal and have equal modules. However, the three-dimensional spin space cannot have more than three such vectors. Consequently, the maximum dimensionality of IR describing equal-module exchange structures does not exceed three in accordance with the results obtained in Refs. [1-2].

## 4. Equal-module exchange phases with coinciding magnetic and crystallochemical unit cells in crystals with spinel structure

Let us consider crystals with a spinel structure by way of illustration of the above discussion. These crystals have the stoichiometric formula $AB_2O_4$, where A and B are cations of metals and O anions of oxygen. Figure 1 shows a primitive cell of a spinel. The crystal lattice symmetry of the spinel is characterized by the space group $O_h^7$. The A and/or B ions can be magnetic ions of 3d elements. These ions occupy crystallographic orbits of the type 8(a) (tetrahedral sublattice) and 16(d) (octahedral sublattice) of the space group $O_h^7$, while anions occupy positions of the type 32(e).

Let us consider commensurate magnetic structures induced by the IR of a space group, which satisfy the Lifshitz condition [11]. For the space group $O_h^7$, the Lifshitz condition is satisfied by 22 IR [6,8,9]. However, the dimensionality of the IR belonging to the star of the wave vector $k_{11} = 0$ does not exceed three. Here and below, we use the same notation for the wave vector stars and IR as in Ref. 8. The dimensionality of the remaining Lifshitz-type IR is higher than three. It is well known [13] that the parameters of magnetic and crystal-lochemical unit cells for exchange structures described by the star of a wave vector equal to zero coincide.

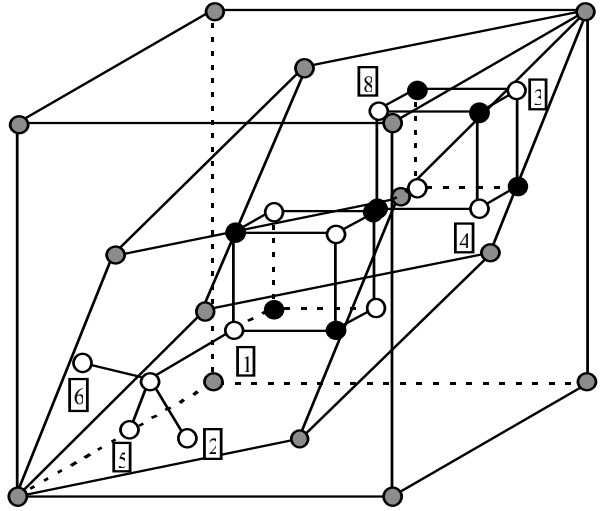

**FIG. 1.** Primitive cell of a spinel crystal. Light circles correspond to 32(e) positions occupied by anions,
dark circles to *16(d)* positions (octahedral sublattice); and hatched circles to 8(a) positions (tetrahedral sublattice).

From the ten IR of the wave vector $\vec{k}_{11}$, the transposition representation includes [5,8] 11-1 and 11-4 IR in positions 8(*a*), 11-1 and 11-7 IR in positions 16(*d*), 11-1 IR being a unit IR according to which magnetic ordering takes place (within a crystallographic orbit). Thus, each sublattice is characterized by a spontaneous magnetization vector. The crystal is ferromagnetic as a whole if the magnetizations of the sublattices are parallel, while the spinel is ferrimagnetic if the magnetizations are antiparallel. Finally, if the magnitudes of magnetization are equal in the latter case, the spinel is a collinear antiferromagnet. Obviously, considerations based on the group theory give no information on mutual orientation of the magnetizations of sublattices, and we must use purely physical arguments concerning the magnitude and sign of intersublattice exchange constants. Let us consider a one-dimensional even IR 11-4 describing an atomic 1:1 ordering in a tetrahedral sublattice. Such an ordering in exchange magnets corresponds to a collinear antiferromagnetic structure. The emergence of the 1:1 ordering lowers the crystal symmetry: $O_h^7 \to T_d^2$. For ordinary phase transitions of the order-disorder type, such a symmetry lowering has the obvious meaning. What is the meaning of such a lowering for a magnetic exchange ordering? The crystal lattice symmetry of a spinel in the paramagnetic and magnetically ordered phases is the same (Fig. 2). This means that neutron diffraction studies of the magnetic structure makes it possible to detect the symmetry group $T_d^2$, while x-ray diffraction analysis gives the symmetry group $O_h^7$ of the crystal lattice. Thus, the transformation of the magnetic symmetry of an exchange-ordered crystal requires two space groups. Exchange ordering according to the 11-4 IR leads to a magnetic structure whose symmetry is described by a binomial symbol $(O_h^7, T_d^2)$.

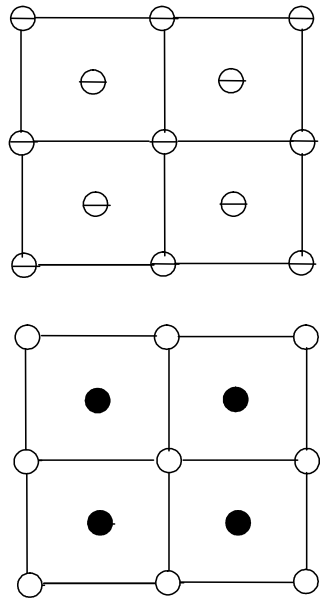

**FIG. 2.** Projection of tetrahedral ions on the *(x,y)* plane in a unit cell of the spinel. Disordered (paramagnetic) phase (a) and ordered (antiferromagnetic) phase (b). The orientation of the average value of spin in positions marked by dark circles is opposite to the orientation of atomic spins occupying positions, marked by light circles. The direction of the antiferromagnetism vector relative to crystallographic axes is arbitrary.

Let us now consider the exchange ordering according to the three-dimensional IR 11-7. Table 1 contains stationary vectors and corresponding symmetry groups for the IR 11-7.

In order to determine the exchange structures induced by the IR 11-7, we supplement Table I with one more table giving the basis functions of this IR in positions 16(*d*). The number 16 in the notation of a crystallographic orbit is the number of points belonging to the given crystallographic orbit and located in a unit cubic cell of the space group $0\backslash$. The primitive cell (see Fig. 1) of this symmetry group contains four nonequivalent positions of the type 16(*d*). In the calculation of basis functions, these four positions are strictly labelled and have the coordinates given in Table 2. The vectors of the primitive cell in Table 2 can be expressed in terms of the vectors of the fcc Bravais unit cell :

$$\vec{a}_1 = \frac{1}{2}(\vec{A}_2 + \vec{A}_3); \vec{a}_2 = \frac{1}{2}(\vec{A}_1 + \vec{A}_3); \vec{a}_3 = \frac{1}{2}(\vec{A}_1 + \vec{A}_2).$$

**TABLE 1.** Exchange structures induced by the IR 11-7 of group

| $\vec{S},\vec{S},\vec{S}$ | $\vec{S},0,0$ | $\vec{S}_1,\vec{S}_2,\vec{S}_2$ | $\vec{S}_1,\vec{S}_2,\vec{S}_3$ |
|---|---|---|---|
| $D_{3d}^5$ | $D_{2h}^{28}$ | $C_{2h}^2$ | $C_i^1$ |

**TABLE 2.** Coordinates of atoms in 16(*d*) position.

| Atom number | Position of atom in primitive cell |
|---|---|
| 1 | $5/8(\vec{a}_1 + \vec{a}_2 + \vec{a}_3)$ |
| 2 | $1/8(\vec{a}_1 + 5\vec{a}_2 + 5\vec{a}_3)$ |
| 3 | $1/8(5\vec{a}_1 + \vec{a}_2 + 5a)$ |
| 4 | $1/8(5\vec{a}_1 + 5\vec{a}_2 + \vec{a}_3)$ |

**TABLE 3.** Scalar basis functions for IR 11-7 for positions 16(*d*)

| Atom number → <br> ↓ Basis functions | 1 | 2 | 3 | 4 |
|---|---|---|---|---|
| $\varphi_1$ | 1 | -1 | -1 | 1 |
| $\varphi_2$ | 1 | -1 | 1 | -1 |
| $\varphi_3$ | 1 | 1 | -1 | -1 |

**TABLE 4.** Magnetic moments of atoms in positions 16(*d*) for antiferromagnetic exchange structures induced by IR 11-7.

| Atom number | Statmatrices | | | |
|---|---|---|---|---|
| | $\vec{S},\vec{S},\vec{S}$ | $\vec{S},0,0$ | $\vec{S}_1,\vec{S}_2,\vec{S}_2$ | $\vec{S}_1,\vec{S}_2,\vec{S}_3$ |
| 1 | $3\vec{S}$ | $\vec{S}$ | $\vec{S}_1 + 2\vec{S}_2$ | $\vec{S}_1 + \vec{S}_2 + \vec{S}_3$ |
| 2 | $-\vec{S}$ | $-\vec{S}$ | $-\vec{S}$ | $-\vec{S}_1 - \vec{S}_2 + \vec{S}_3$ |
| 3 | $-\vec{S}$ | $-\vec{S}$ | $\vec{S}$ | $-\vec{S}_1 + \vec{S}_2 - \vec{S}_3$ |
| 4 | $-\vec{S}$ | $\vec{S}$ | $\vec{S}_1 - 2\vec{S}_2$ | $\vec{S}_1 - \vec{S}_2 - \vec{S}_3$ |

We can now obtain scalar basis functions calculated for these four atoms [13] (Table 3).

Substituting the basis functions from Table 3 and the elements of statmatrix from Table I into formula (1), we obtain the average value of the spin moment for each magnetic atom in the position 16(*d*) (Table 4). Table 4 shows that the total magnetic moment of a primitive cell is equal to zero for all the four phases as it should be in the case of the antiferromagnetic ordering. One-parametric phases (S,S,S) and (S,0,0) correspond to collinear antiferromagnetic structures, the latter phase corresponding to an Andreev-Marchenko equal-module exchange structure. The two-parametric phase $(S_1, S_2, S_2)$ corresponds to a two-dimensional antiferromagnetic structure. The three-parametric phase with the lowest symmetry corresponds to an essentially three-dimensional antiferromagnetic structure. If the three vectors $\vec{S}_1, \vec{S}_2, \vec{S}_3$ have equal magnitudes and are mutually orthogonal, they satisfy conditions (8) and (9), and we again have a three-dimensional equal-module antiferromagnetic structure. Under certain thermodynamic conditions, three vectors $\vec{S}_1, \vec{S}_2, \vec{S}_3$ can accidentally become equal in magnitude and mutually orthogonal, but these conditions do not correspond to any thermodynamic phase since the conditions (7) and (8) for the existence of an equal-module structure are not thermodynamic conditions. Conversely, the equal-module structure corresponding to the one-parametric solution (S,0,0) is obtained under certain thermodynamic conditions and corresponds to a stable (in certain limits) equal-module exchange structure.

**5. Conclusion**

We can summarize the results of the above analysis as follows. Equal-module exchange structures that can be formed in crystals belong to two essentially different classes. The first class includes thermodynamically stable exchange structures characterized by their own magnetic symmetry. The second class includes the structures appearing under conditions (7) and (8). These structures can be singled out from a thermodynamically stable exchange structure by imposing the additional nonthermodynamic condition (8) and have no magnetic phase with its own magnetic symmetry corresponding to them. The problem of exchange structures is not only of academic interest and does not serve just as an exercise in the methods of the group theory. The results of numerous experiments show that magnetic phases (see Refs. [14 – 15]) whose thermodynamic properties are determined only by the exchange interaction can be formed in some magnetic crystals. For this reason, the symmetry of exchange structures exactly corresponds to the symmetry of isotropic magnetic phases in these crystals.

———————————————

*Electronic address: borlakov@mail.ru

1. Andreev A F, Marchenko V I //*Sov. Phys. Usp.* **23** 21–34 (1980).
2. V. G. Bar'yakhtar and D. A. Yablonskii, Fir. Nizk. Temp. **6**, 345 (1980) [Sov. J. Low Temp. Phys. **6**, 164 (1980)].
3. Yu. M. Gufan, E. I. Kul'in, V. L. Lorman at *al.*//JETP Lettt. **46**, 287 (1987).
4. Yu. M. Gufan, *Structural Phase Transitions* [in Russian], Nauka, Moscow (1982).
5. V. P. Sakhnenko, V. M. Talanov, and G. M. Chechin. Fiz. Mel. Melalloved. 62, 847 (1986).
6. V. P. Sakhnenko, V. M. Talanov, and G. M. Chechin, Dep. V1NITI, dep. No. 638-82, Moscow (1982).
7. Kh. Sh. Borlakov, Generalization of equal-module exchange magnetic classes// LOW TEMP. PHYS. -1998.-Vol.24, no. 9-pp.647-651.
8. Izyumov, Yu.A., Naish, V.E., and Ozerov, R.P.,  Neutron Diffraction of Magnetic Materials; Consultants Bureau: New York, NY, USA, 1991.
9. O. V. Kovalev, Representations of the Crystallographic Space Groups: Irreducible Representations, Induced Representations, and Corepresentations, Yverdon, Switzerland ; Philadelphia, Pa., U.S.A. : Gordon and Breach Science, 1993.
10. N. N. Bogoliubov and D. V. Shirkov, Introduction to the Theory of Quantized Fields Wiley, New York,  (1980).
11. Landau L.D. and Lifshits E.M.: Statistical Physics, Oxford:Pergamon, (1980) 1, 3rd edition.
12. V. L. Indenbom. Sov. Phys. Crystallogr. **5**, 115 (I960).
13. V. P. Sakhnenko, V. M. Talanov, and G. M. Chechin, Dep. VINITI, dep. No. 6379-83, Moscow (1983).
14. Kh. Sh. Borlakov, //Phys of Metalls and Metallography.-1999.-Vol.88,No. 1-pp.19-27
15. Kh. Sh. Borlakov, //Crystall. Rep.-2001.-Vol.46,No. 1-pp.88-91.